\documentclass[prd, 11pt, a4paper, showpacs, notitlepage, dvipsnames, reprint, twocolumn, preprintnumbers, nofootinbib]{revtex4-1}

\usepackage{geometry}
\usepackage[utf8x]{inputenc}
\usepackage{makecell}
\usepackage{float}
\usepackage{tabularx}
\usepackage{enumitem}
\usepackage{ulem}
\usepackage[activate={true,nocompatibility},final,tracking=true,kerning=true,factor=1500,stretch=10,shrink=10]{microtype}
\usepackage{graphicx}
\usepackage{bm, amssymb, amsmath, amsfonts}
\usepackage{multirow}
\usepackage{xcolor}

\usepackage{url}

\usepackage{url}
\usepackage[breaklinks]{hyperref}

\usepackage{hyperref}
\hypersetup{
colorlinks = true, 	    
linkcolor = magenta,	
citecolor = blue,		
}

\usepackage[linesnumbered,ruled,lined,boxed]{algorithm2e}
\usepackage[noend]{algpseudocode}

\SetAlCapSty{xAlCapSty}

\SetCommentSty{mycommfont}

\SetNlSty{mynlfont}{}{} 

\RestyleAlgo{algoruled}

\usepackage{subcaption}

\setcitestyle{numbers,square}

\begin{document}

\preprint{LIGO DCC number \bf LIGO-P2200119}

\title{
Convolutional neural network for gravitational-wave early alert: \\ Going down in frequency
}

\author{Grégory Baltus$^{1}$}	\email[]{gbaltus@uliege.be}
\author{Justin Janquart$^{2,3}$} \email[]{j.janquart@uu.nl}
\author{Melissa Lopez$^{2,3}$}	\email[]{m.lopez@uu.nl}
\author{Harsh Narola$^{2,3}$} \email[]{h.b.narola@uu.nl}
\author{Jean-René Cudell$^{1}$} \email[]{jr.cudell@uliege.be}

\affiliation{${}^1$ STAR Institute, Bâtiment B5, Université de Liège, Sart Tilman B4000 Liège, Belgium}
\affiliation{${}^2$ Nikhef, Science Park 105, 1098 XG Amsterdam, The Netherlands}
\affiliation{${}^3$ Institute for Gravitational and Subatomic Physics (GRASP), Utrecht University, Princetonplein 1, 3584 CC Utrecht, The Netherlands}

\begin{abstract}
We present here the latest development of a machine-learning pipeline for pre-merger alerts from gravitational waves coming from binary neutron stars. This work starts from the convolutional neural networks introduced in \cite{baltus2021convolutional} that searched for three classes of early inspirals in simulated Gaussian noise colored with the design-sensitivity power-spectral density of LIGO. Our new network is able to search for any type of binary neutron stars, it can take into account all the detectors available, and it can see the events even earlier than the previous one. We study the performance of our method in three different types of noise: Gaussian O3 noise, real O3 noise, and predicted O4 noise. We show that our network performs almost as well in non-Gaussian noise as in Gaussian noise: our method is robust w.r.t. glitches and artifacts present in real noise. Although it would not have been able to trigger on the BNSs detected during O3 because their signal-to-noise ratio was too weak, we expect our network to find around 3 BNSs during O4 with a time before the merger between 3 and 88 s in advance.
\end{abstract}
\maketitle

\section{Introduction}
\label{Sec:Intro}
Multi-messenger astrophysics (MMA) makes use of {messengers from different forces} of the Universe to  provide a wealth of information about various astrophysical processes. From previous investigations it is well-known that the combination of at least two of these signals gives qualitatively different and complementary types of information, capable of probing down to the densest and {most energetic} regions of cosmic objects, which were hidden from astronomers’ sight up until now \cite{Meszaros:2019xej, SN1987A, MAGIC:2018sak}.

In the context of gravitational waves (GW) {combined} with other astrophysical signals, it has long been suggested that short gamma ray burst (GRB) might be related to binary neutron star (BNS) mergers \cite{goodman1986gamma}, which fall in the sensitivity band of second generation ground based-detectors \cite{LIGOScientific:2014pky, VIRGO:2014yos, KAGRA:2018plz}. Several studies investigated the expectations of electromagnetic (EM) follow-up efforts during the Advanced LIGO and Virgo era of compact binary coalescence (CBC) \cite{Cannon:2011vi, Singer_2014}.
On August 17, 2017 the Fermi Gamma-ray Burst Monitor \cite{meegan2009fermi} announced the detection of a GRB, later designated as GRB170817A \cite{Goldstein_2017}. Approximately 6 minutes later, a GW candidate was registered in low latency based on a single-detector analysis of the Advanced Laser Interferometer Gravitational-wave Observatory (LIGO) Hanford data, which after a rapid re-analysis of data from LIGO and Virgo, it was re-labeled as GW170817 \cite{LIGOScientific:2017ync}. An extensive observing campaign was launched across the electromagnetic spectrum in response to the Fermi-GBM and LIGO–Virgo triggers, which led to the detection of the kilonova associated with  GW170817, later called AT 2017gfo \cite{Perego:2017wtu}. 

In recent times, there has been a sparkling interest in early warning (or pre-merger) alerts of BNS in the field of GW for EM and astro-particle follow-ups \cite{Sachdev:2020lfd, baltus2021convolutional, Yu:2021vvm, tsutsui2021early, nitz2020gravitational, nitz2018rapid, sachdev2020early}, since the radiation emitted from these systems enters the sensitive region of the interferometers during the inspiral phase \cite{10.1093/astrogeo/atab077, Magee:2021xdx}.
The predicted rates for these joint detections are $0.02−27$ per year for X-ray band, 0.01−19 per year for optical band, and 0.02−25 per year for radio band, respectively, at design sensitivity for a three detector network \cite{Saleem:2017pvc}. It is relevant to note that the large uncertainty is due to the fact that BNS merger rate is not well constrained. 
The improving sensitivity of second-generation detectors and the even better sensitivity for the third-generation detectors, such as Cosmic Explorer and the Einstein Telescope~\cite{Cosmic_Explorer, Einstein_telescope} will lead, via an increase in the signal-to-noise ratio (SNR), to a major improvement in the early-detection and sky-localization capabilities \cite{nitz2021pre}.
Another key element to develop MMA further is the design of low-latency pipelines for the production of real-time GW alerts, or even pre-merger alerts. The current state-of-the-art employs matched filtering techniques \cite{Cannon:2011vi} to perform online analyses via the pipelines GstLAL~\cite{sachdev2019gstlal}, PyCBCLive 
~\cite{nitz2018rapid}, MBTAOnline ~\cite{adams2016low} and SPIIR
~\cite{chu2022spiir}. We refer the reader to ~\cite{LIGOScientific:2019gag} and ~\cite{abbott2021gwtc_2p1, LIGOScientific:2021djp, davis2021ligo} for a summary of the low-latency efforts carried out by the LIGO-Virgo collaboration during the second and third observing runs. 

Recent investigations in the GW field have focused on Machine Learning (ML) algorithms, due to their success in different tasks and domains. The main advantage of ML techniques is their rapidity because most of the computations are made during the training stage.
A widely used ML method for pattern recognition is based on convolutional neural networks {(CNNs)} \cite{Goodfellow-et-al-2016}, in the context of GW it has been applied to different tasks such as  CBC identification \cite{Gabbard:2017lja, gebhard2017convwave, Gebhard:2019ldz, KRASTEV2020135330, schafer2020detection}, burst detection \cite{Astone:2018uge, Iess:2020, LopezPortilla:2020odz, Boudart:2022xib}, sky localization \cite{Green:2020hst, Williams:2021qyt, Kolmus:2021buf}, glitch classification \cite{Zevin:2016qwy, Soni:2021cjy} and synthetic data generation \cite{McGinn:2021jqg, Lopez:2022lkd}. See \cite{Cuoco:2020ogp} for a review on this topic. 

ML methods have also emerged as a new tool  in the context of early warning \cite{baltus2021convolutional, Yu:2021vvm}, allowing us to flag prompt triggers for GW candidates. The final goal of this work is to detect BNS signals before the merger. To do that, we have design a single CNN that takes as input the time-series data from all the online detectors and returns a classification between two classes: pure noise or noise plus inspiral. In this paper, we build on our previous work \cite{baltus2021convolutional},
improving on the techniques previously developed, and testing them on more realistic scenarios:  we use real O3 noise, as well as the data from all available detectors. In addition, we retrained our network on predicted O4 noise and give expected efficiencies for this run. 

The details of the differences with \cite{baltus2021convolutional} are as follows:
\begin{itemize}
    \item[--] the addition of the spin effect to the BNS waveforms; 
    \item[--] a uniform sky location of the injections;
    \item[--] the injection of simulated BNS signals in simulated O3 noise, real O3 noise and simulated O4 noise;
    \item[--] a decrease of the minimal cutoff frequency from 20 Hz to 10 -- 15 Hz;
    \item[--] a fixed input-signal duration of 300 s with a sampling frequency of 512 Hz that allows to analyse any BNS signal for all allowed neutron star masses; 
    \item[--] the implementation of curriculum learning \cite{bengio2009curriculum}.
\end{itemize}

This paper is organized as follows: in Section \ref{Sec:Method}, the method is explained. Subsection \ref{Subsec:PISNR} introduces the definition of the SNR and the partial inspiral signal-to-noise ratio (PISNR) used in this work, as well as the relation between the frequency of a waveform and the time before the merger. The description of the data generation and the training strategy is made in Subsection \ref{Subsec:data}. The last part of this section, \ref{Subsec:Network},  describes the architecture of the CNN used in this paper. Section \ref{Sec:Results} presents the results and the performance of our method in the three types of noise, as well as studying the number of BNS that are expected to be found in advance by our network in O4. Finally, we give our conclusions in section \ref{Sec:Conclusion}.

\section{Method}
\label{Sec:Method}
\subsection{Partial-Inspiral Signal to Noise Ratio}
\label{Subsec:PISNR}
In GW-searches, the matched-filtering SNR ($\rho$)~\cite{nitz2018rapid} is used to verify how well a template matches the data. The SNR definition follows that of the FINDCHIRP algorithm \cite{PhysRevD.85.122006}
as implemented in PyCBC \cite{Biwer:2018osg}. One first transforms
the signal $s(t)$ and templates $h(t)$ to frequency space:
\begin{equation}
\tilde{h}(f)=\int_{-\infty}^{\infty}h(t)e{}^{-2\pi ift}dt
\end{equation}
and similarly for $\tilde{s}(f)$. One can then define a complex matched filter
\begin{equation}
z(t)=4\int_{0}^{\infty}\frac{\tilde{s}(f)^{*}\tilde{h}(f)}{S_{n}(f)}e^{2\pi ft}df
\end{equation}
where $S_{n}(f)$ is the one-sided noise strain power spectral density (PSD)
of the detector and the * superscript denotes
complex conjugation. It can be shown \cite{Anderson:2007km} that, after
minimising w.r.t the phase of the signal at the time of entry in the
frequency band of the interferometer, the standard matched filter
expression can be written $|z(t)|.$ It can also be shown that its
variance is given by 
\begin{equation}
\sigma^{2}=4\int_{0}^{\infty}\frac{\tilde{h}(f)^{*}\tilde{h}(f)}{S_{n}(f)}df.
\end{equation}
The signal-to-noise ratio is then taken to be

\begin{equation}
\label{eq:SNR}
\rho(t)=\frac{|z(t)]}{\sigma}.
\end{equation}

For a network of $N$ detectors, identified by an index $i=1...N$,
one defines the network SNR as
\begin{equation}
\rho_{net}(t)=\sqrt{\sum_{1}^{N}\rho_{i}^{2}(t)}.
\end{equation}

The SNR is a key quantity for the searches based on matched filtering, since it describes the amount of overlap between a template and an unknown signal. In these searches, the strategy is to create a template bank of pre-computed waveforms and use it to calculate the SNR over all the data strain. As a first step,  a trigger is created when the SNR reaches a maximum value higher than a given threshold. After that, it undergoes a statistical treatment to be confirmed as a GW candidate \cite{nitz2018rapid}. 
As we are interested in searching for the early inspiral, a more meaningful indicator will be the partial-inspiral signal-to-noise ratio. It is defined as the SNR in which the template $h$ is the partial template that contains only the fraction of the inspiral part of the waveform that our network tries to identify. For more details about the PISNR, and how it evolves depending on the length of the template, we refer to section II. A. of Ref.~\cite{baltus2021convolutional}.

\begin{figure}[ht]
\begin{center}
    \includegraphics[scale=0.5]{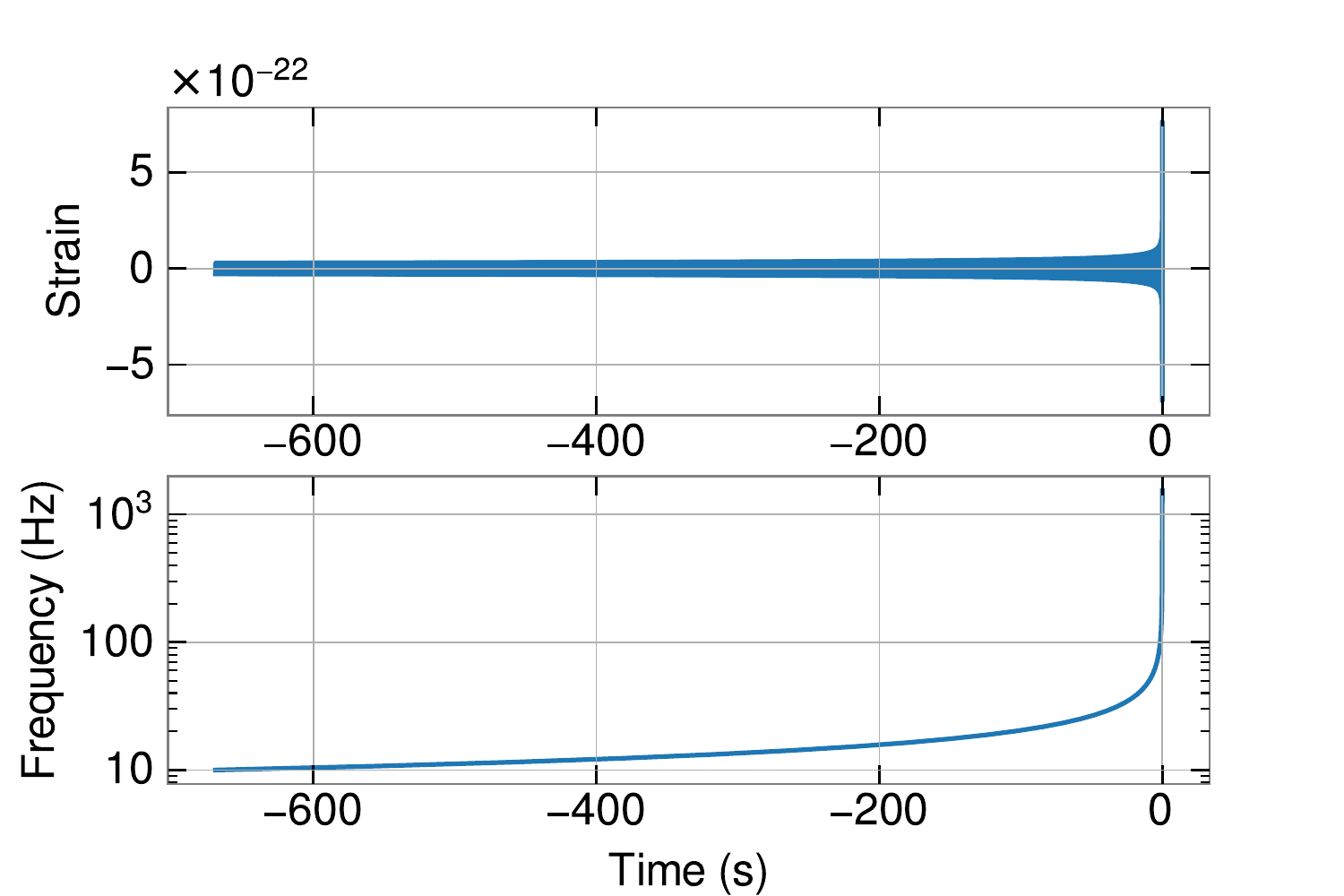}
    \caption{\label{fig:FreqEvolution} The top figure represents a GW signal corresponding to two objects of mass 1.8 $M_\odot$. The bottom figure represents the evolution in frequency  for this binary.} 
\end{center}
\end{figure}

At the lowest order in velocity, the frequency $f$ at a time $t$ depends on the chirp mass $\mathcal{M}_c$ of the system and the merger time $t_{m}$:
\begin{equation}
\label{eq:ft_deltat}
    f(t) = \frac{1}{\pi} \bigg(\frac{G \mathcal{M}_{c}}{c^{3}}\bigg)^{-5/8} \bigg(\frac{5}{251}\frac{1}{(t_{m}-t)}\bigg)^{3/8}.
\end{equation}
So for a given chirp mass, if we say that we can detect an event $(t_{m}-t)$ seconds before the merger, it is equivalent to say that we detect the signal when the maximum frequency is $f(t)$.
Fig.~\ref{fig:FreqEvolution} represents the time and frequency evolution for a GW.

\subsection{Data and training strategies}
\label{Subsec:data}
\begin{figure}[ht]
\begin{center}
    \includegraphics[scale=0.5]{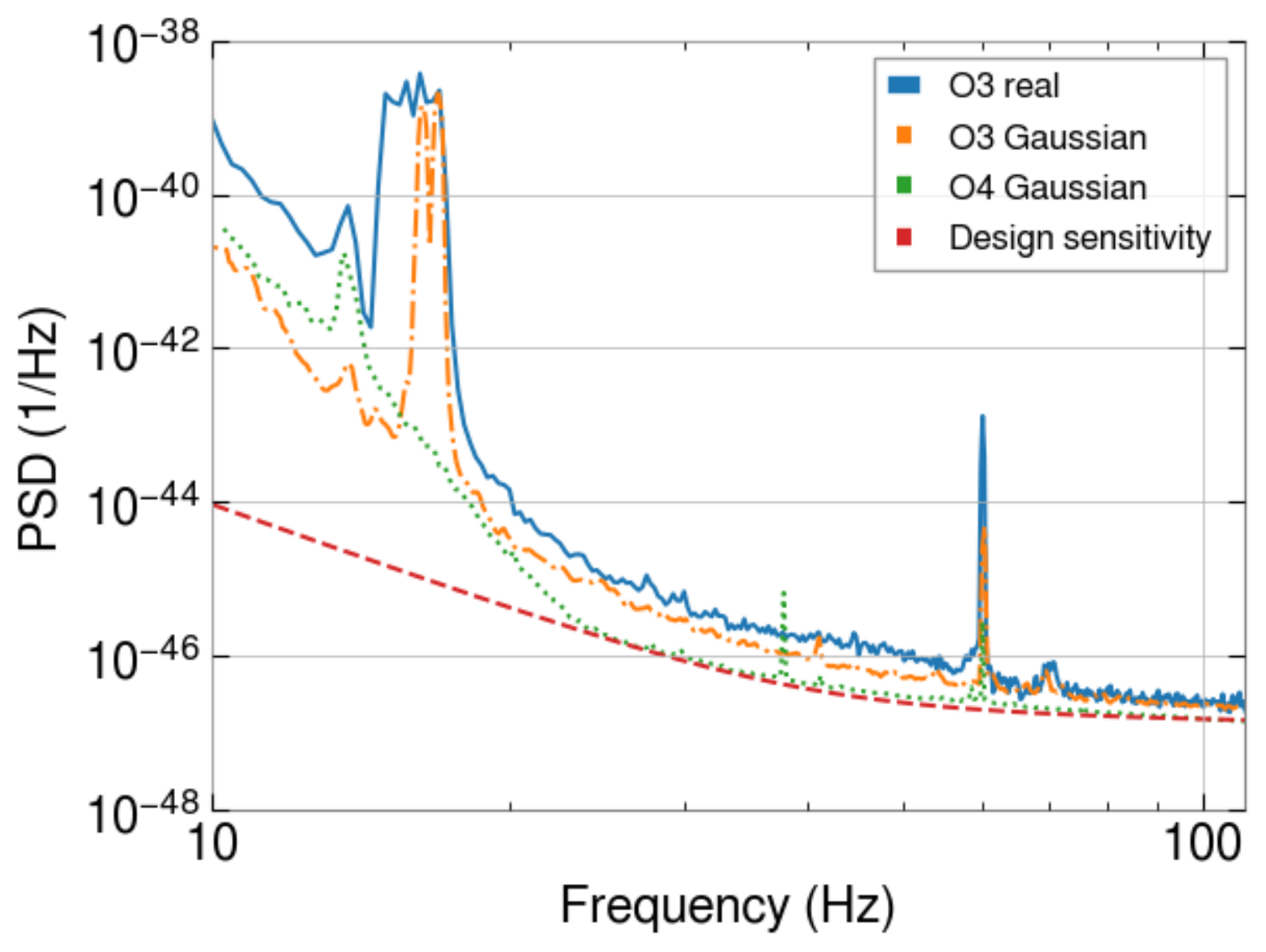}
    \caption{\label{fig:PSDs} Representation of the different PSDs, for the Livingston detector, used to generate the different data sets. We also show the design sensitivity PSD provided by PyCBC~\cite{Biwer:2018osg} used in~\cite{baltus2021convolutional}.} 
    \label{PSDs}
\end{center}
\end{figure}

Three different types of noise were considered in this paper: 
\begin{enumerate}[label=\itshape\alph*)]
    \item O3 Gaussian noise.
    \item real low-latency O3 noise.
    \item O4 Gaussian noise.
\end{enumerate}
The corresponding PSDs are represented in Fig. \ref{PSDs}. To generate a frame of simulated O3 Gaussian noise, we use the theoretical PSDs provided by PyCBC~\cite{Biwer:2018osg}\footnote{The PSD used for Gaussian O3 LIGO is \textit{aLIGOaLIGO140MpcT1800545}, the one for Virgo is \textit{aLIGOAdVO3LowT1800545}, both coming from PyCBC~\cite{Biwer:2018osg}.}. To obtain data of O3, we directly download the low-latency strain of the detectors~\cite{abbott2021gwtc_2p1, LIGOScientific:2021djp, davis2021ligo} using the \textit{GWpy} package~\cite{gwpy} \footnote{To download the real O3 low-latency data, we use the channels H1:GDS-CALIB\_STRAIN, L1:GDS-CALIB\_STRAIN, V1:Hrec\_hoft\_16384Hz, and the frame type: H1\_llhoft, L1\_llhoft, V1Online in \textit{GWpy}. }. To generate the O4 Gaussian noise, we use the predicted O4 PSD coming from the observing scenarios \cite{abbott:ObservingScenario,petrov:ObservingScenario}\footnote{The LIGO and Virgo PSDs used for O4 correspond to the ones shown in Fig.~1 of~\cite{abbott:ObservingScenario}, with the BNS detector horizon at 160 Mpc for the LIGO detectors and the horizon at 120 Mpc for the Virgo detector.}.

\begin{figure}[ht]
\begin{center}
    \includegraphics[scale=0.5]{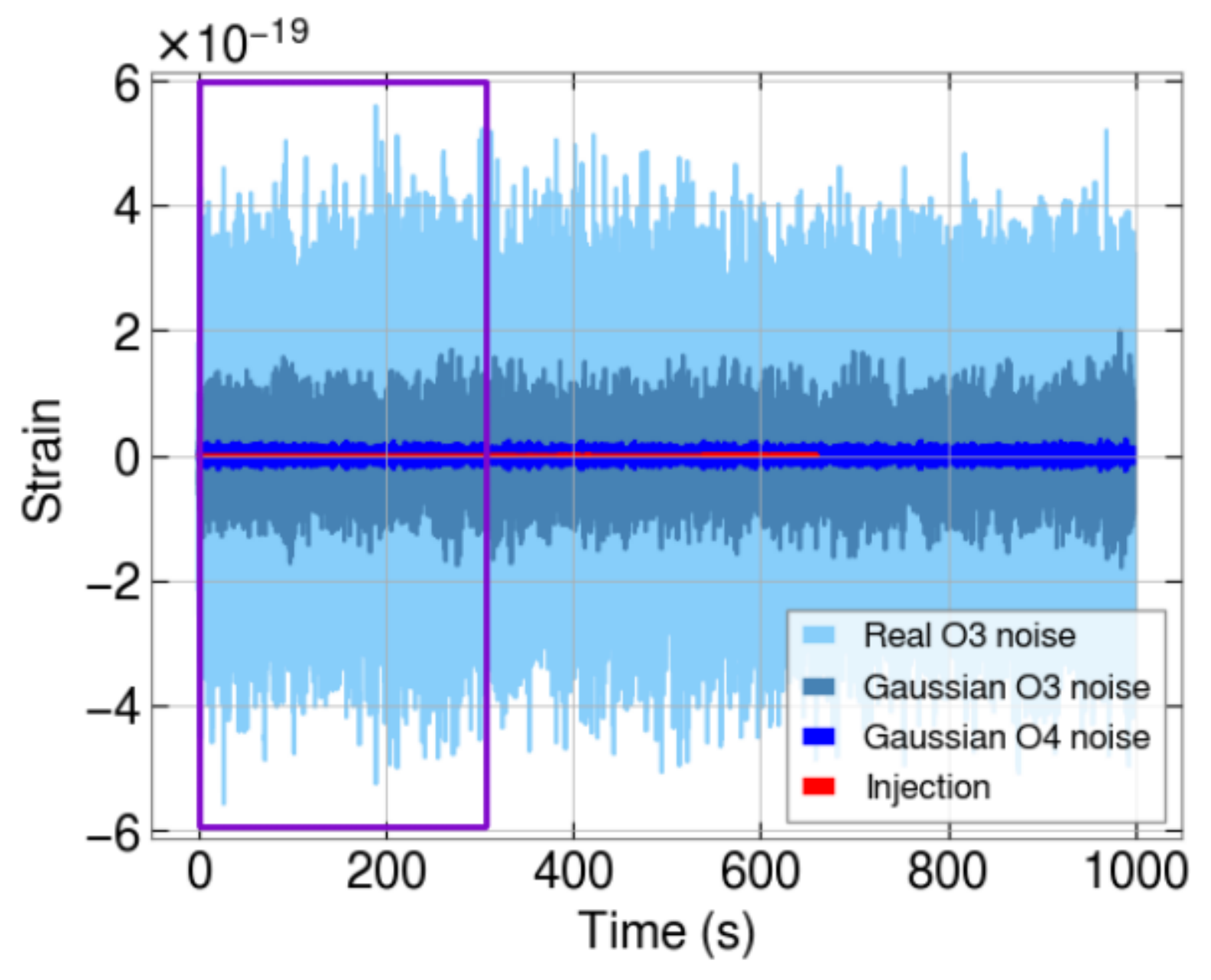}
    \caption{\label{fig:InjNoise} Representation of the different types of noises for the Handford detector used together with an injection similar to GW170817, i.e. with neutron star masses of $1.46$ and $1.27$ $M_\odot$ \cite{abbott2017gw170817}. Note that only the part in the rectangle is passed to the network.} 
\end{center}
\end{figure}

Since the problem at hand can be solved as a classification task, we need a data set containing two classes: noise and noise plus inspiral, also known as injections. For the injections, we generate waveforms using the approximant \textit{SpinTaylorT4} \cite{Buonanno:2002fy}. 
We choose the component masses to be uniformly distributed between 1 and 3 solar masses to cover all the possible BNS systems \cite{kiziltan2013neutron}. The sources are uniformly distributed over the sky, and we also include the spin effects. With these parameters, the simulated signal is always longer than 300 seconds. In such a way, the inputs of the network contain only the early inspiral part (see Fig.~\ref{fig:InjNoise} for an illustration).
After injecting the simulated signal into the noise the frames are whitened, and we apply a low-pass filter at 100 Hz and a high-pass filter at 10 Hz. For O3 real noise, some significant peaks can appear in the whitened strain due to non-Gaussian effects. In our approach, these effects are vetoed by zeroing them out, see appendix \ref{appendix:1}. Afterwards, the final frame is renormalized, making all the values in the frame between $-1$ and $1$.

For the training and testing, we choose a distribution in distance such that the distribution in PISNR is an inverse Gaussian with a mean of thirty-five and a scale of one hundred \footnote{We found that the inverse Gaussian (Wald) distribution fits better our goal. Indeed, this distribution gives a few very high PISNR events that enable the network to start its learning process.}. Despite having a large data set containing one million, we have observed a low performance when we decrease the maximum frequency to $\sim 25$ Hz. This is because the CNN is good to detect a variation of frequency, and for earlier inspiral phase, the signal becomes more monochromatic.  

To be able to detect events earlier, it is key to decrease the maximum frequency seen by our model. For this aim, we change the training strategy and use curriculum learning \cite{bengio2009curriculum} as a function of the maximal frequency seen by the network, as it has shown an increase of the performance as a function of the SNR in previous works \cite{LopezPortilla:2020odz}. The principle of curriculum learning is to train the network first on easier data (on data with a high maximum frequency), and then gradually increasing the difficulty (on data with a lower maximum frequency).  The network is then iteratively trained on each training set. To prevent the network from forgetting what it has learned, we keep all the data of the previous steps while adding the new ones. To that effect, we generate five different training sets. The parameter distributions for the injections stay the same, except for the maximal frequency seen by the network. This parameter is now chosen as a Gaussian distribution with a standard deviation of 2.5 Hz, and different mean depending on the data sets. More information about these data sets can be found in table~\ref{tab:MinMaxFreq}. Each training set contains 20000 frames and half of them contain an injection. Note that 20 \% of each training set was used for validation during the training. 
For each step, we train for six epochs as it was enough to make the loss converge without facing over-fitting. The use of curriculum learning allows to improve the performance on data set 3, 4, 5 with maximum frequency of, respectively, $30$, $25$, $20$ Hz, while maintaining the performance at higher frequencies. We did not create a data set with a maximum frequency smaller than $20$ Hz, because the sensitivity of the detector is poor under this value, see Fig.~\ref{fig:PSDs}. 

\begin{table}[t!]
    \begin{tabular}{ |c|c|c|c|c|c|  }
     \hline
     Data set & Max Freq & Min Freq & Min TBM & Max TBM\\
     \hline
     Data set 1 & 40 Hz & 12.9 Hz & 7 s  & 44 s\\
     Data set 2 & 35 Hz & 12.8 Hz & 10 s & 63 s\\
     Data set 3 & 30 Hz & 12.6 Hz & 15 s & 95 s\\
     Data set 4 & 25 Hz & 12.3 Hz & 24 s & 115 s\\
     Data set 5 & 20 Hz & 11.7 Hz & 45 s & 280 s\\
     \hline
    \end{tabular}
    \caption{\label{tab:MinMaxFreq} Each data set corresponds to a value for the maximum frequency seen by the networks, which in turn leads to a minimum frequency and a Time Before the Merger (TBM). The numbers shown for the maximum and minimum frequency are the mean value in each data set. The maximum and minimum TBMs are the TBM for two objects of respectively 3 $M_\odot$ and 1 $M_\odot$. 
    }
\end{table}

The training on the real noise data was done in a similar way. Note that we have done the training with noise coming only from O3a, meaning the first half of O3 \cite{abbott2021gwtc_2p1}. We have vetoed the time of the real events from the GWTC-2.1 catalog \cite{abbott2021gwtc_2p1} not to train on them, as most of them were BBH. For all the testing, we used low-latency noise coming from O3b, the second half of O3 \cite{LIGOScientific:2021djp, davis2021ligo}. 
During O3 there are times when not all detectors are online. To take this fact into account, when a certain detector is offline, we feed the corresponding channel with a vector of zeros. In this way, our network is able to perform the search regardless of the number of detectors available. 

For the training parameters, we use a batch size of 50. For each step of curriculum learning we train for 6 epochs, it was enough to make the loss converge. The learning rate is $8 \times 10^{-5}$ and the optimizer is ADAMAX with a weight decay of $10^{-5}$. ADAMAX is a variant of ADAM, based on the infinity norm~\cite{kingma2019method}.

We use the weighted cross-entropy loss \cite{de2005tutorial}. At first, we employed the cross-entropy loss, which is standard for classification problems. However, this led to a large number of false positives. To remedy that we decided to weigh this loss \cite{LopezPortilla:2020odz} by a factor 0.4 for the frames with an injection. This  reduces the chances that the network classifies a frame containing only noise as an event, so it reduces the number of false positives. We tried multiple values for the weight and found that for the task at hand a factor of 0.4 translates into a reduction of the FAR while maintaining the TAP performance.
\subsection{Description of the network}
\label{Subsec:Network}
\begin{figure*}[ht]
\begin{center}
    \includegraphics[scale=0.4]{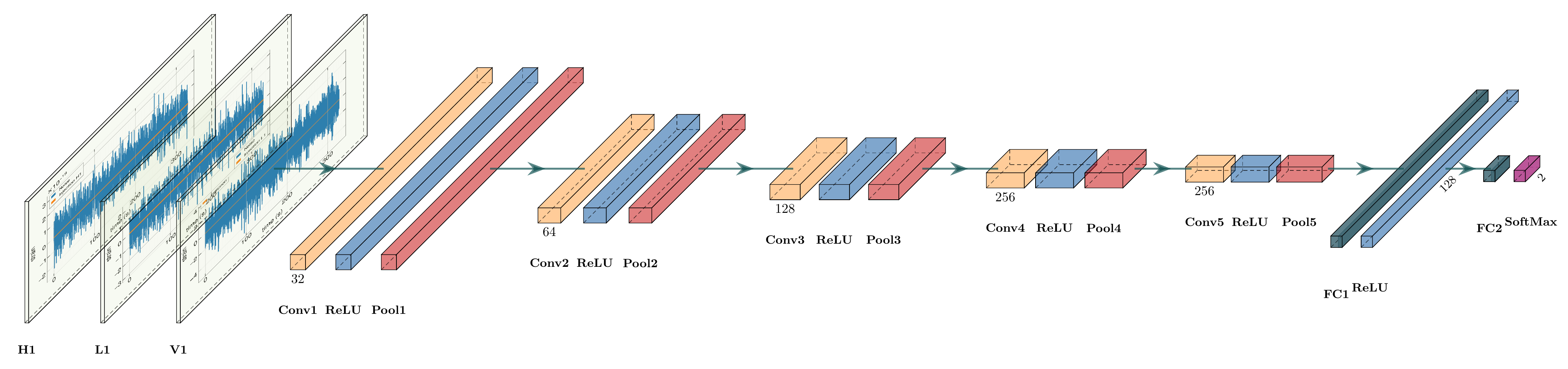}
    \caption{\label{fig:CNN} Representation of the CNN architecture, the yellow layers are the convolutions, the blue ones are the ReLU activation, the red ones are the pool layers, the purple ones are the dense layers, and the dark purple is the final softmax layer. The number under each layer represents the number of channels.} 
\end{center}
\end{figure*}

The architecture of the network is similar to that in Ref.~\cite{baltus2021convolutional}. A representation of the neural network is given in Fig.~\ref{fig:CNN}, we use the \textit{Pytorch} package to create the architecture \cite{NEURIPS2019_9015}. The network takes 300 seconds of data for each available detector. In other words, it has three input channels, each corresponding to one of the three detectors (Hanford, Livingston, and Virgo)\footnote{The Conv1D layer as implemented in PyTorch allow us to give as input any number of channels, see \url{https://pytorch.org/docs/stable/generated/torch.nn.Conv1d.html} \cite{NEURIPS2019_9015}}. It is composed of a batch normalisation layer, followed by 5 blocks composed of a convolution layer, a ReLU activation, and a pool layer. For the convolution, the kernel sizes are successively 16, 8, 4, 8, 16. For the pool layers, the kernel size is always set to 4. The stride is set to 1 for the convolution layers and 4 for the pool layers. After these blocks, we add two linear layers with sizes of respectively 128 and 2 interspersed by a ReLU activation. The final layer is a softmax layer that returns a probability vector. 

\section{Results}
\label{Sec:Results}
\subsection{Performance of the network}
\label{Subsec:Performance}
After the training, the testing sets come from the same distribution as the training sets, see table~\ref{tab:MinMaxFreq}. The other parameter distributions are the same as for the training sets. Each of the test sets contains 4400 frames, half of which are pure noise and half noise plus injection. The total size of the test sets for a type of noise is then 22000 frames.

\begin{figure}[ht]
\begin{center}
    \includegraphics[scale=0.4]{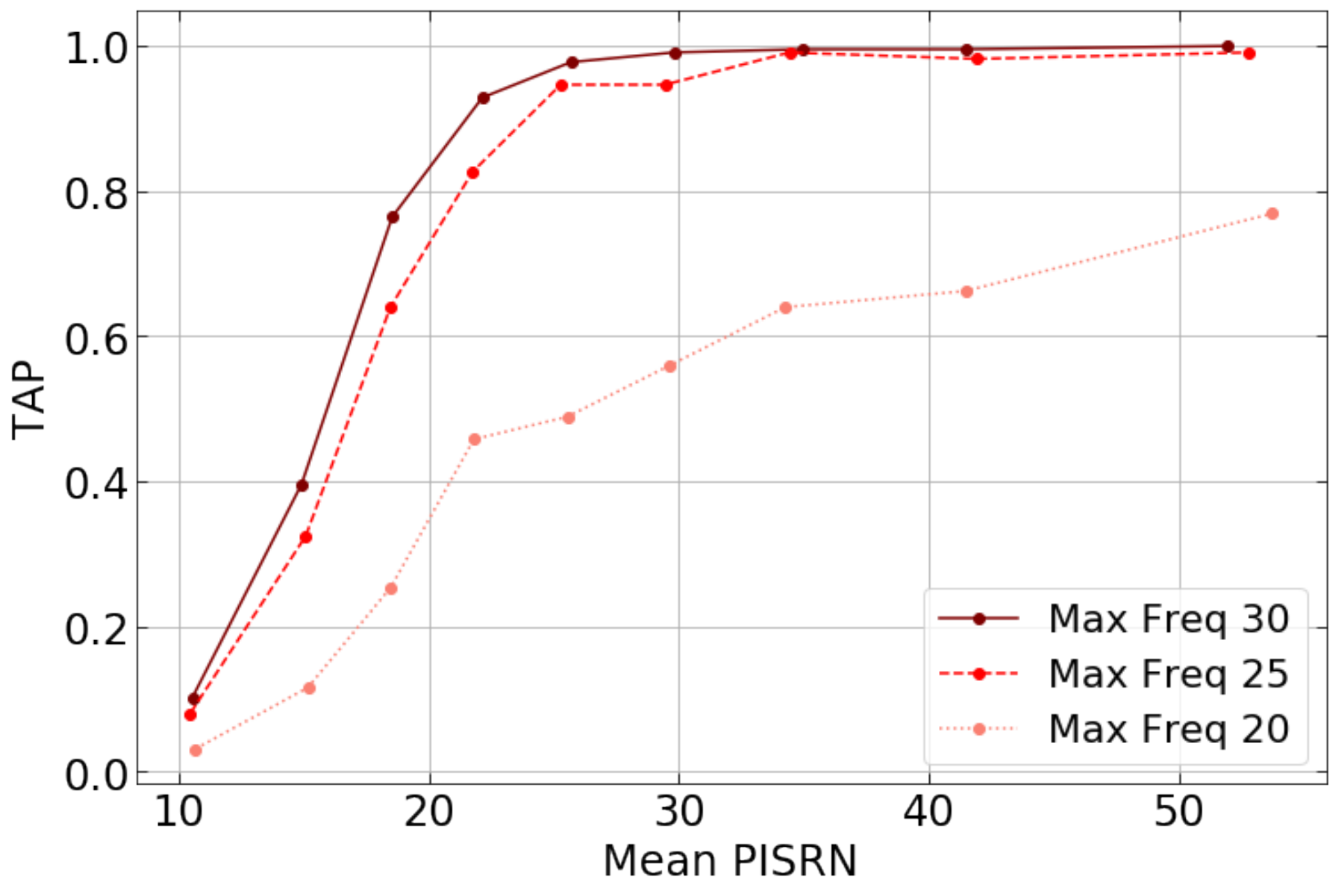}
    \caption{\label{fig:TAPO3gaussian}The True Alarm Probability as a function of the PISRN for the O3 Gaussian noise case. Each curve represents a different test set with a different maximum frequency seen by the CNN. } 
\end{center}
\end{figure}

The efficiency of our network for the different steps of curriculum learning can be seen in Fig.~\ref{fig:TAPO3gaussian}. We define the True Alarm Probability (TAP) and the False Alarm Probability (FAP) as equation 7 in our previous work \cite{baltus2021convolutional}.
In Fig.~\ref{fig:TAPO3gaussian}, we represent the three lowest maximum frequencies data set of curriculum learning, as the higher maximum frequencies have performances similar to the $30$ Hz data set. For the data sets with a maximum frequency $> 25$ Hz, an efficiency of $50\%$ is obtained at $\sim 15$ PISNR, while the efficiency reaches $100\%$ at $30$ PISNR. This is not the case for the data set with a maximum frequency of $20$~Hz, where the TAP is lower. This is expected since the sensitivity of the detectors becomes worse at lower frequencies, typically under 20~Hz, see Fig.~\ref{fig:PSDs}. In all the figures shown in this work, the FAP is fixed at $1\%$.

\begin{figure}[ht]
\begin{center}
    \includegraphics[scale=0.4]{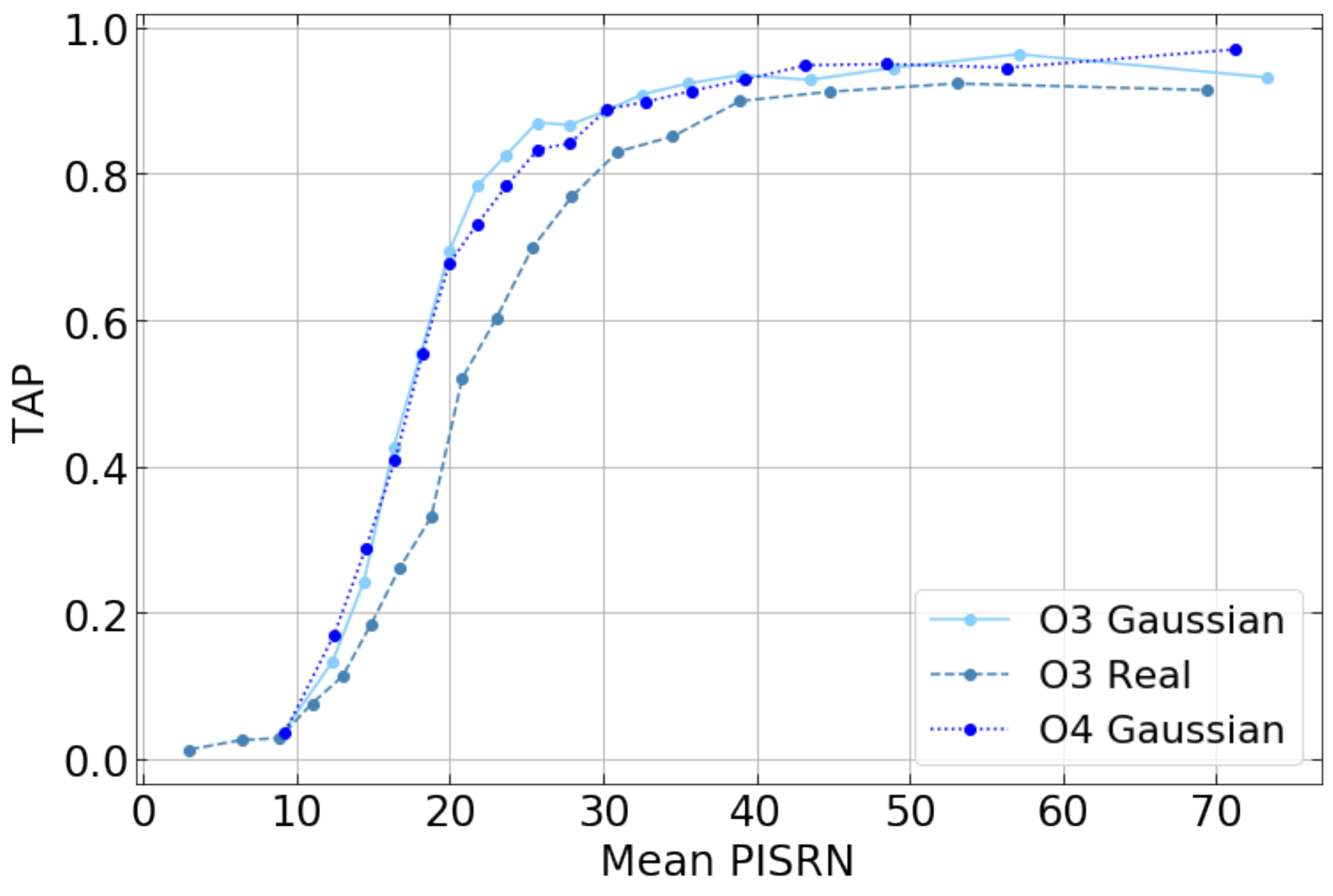}
    \caption{\label{fig:TAPcomp} The True Alarm Probability as a function of the PISRN for the O3 Gaussian noise, real O3 noise, and O4 Gaussian noise.} 
\end{center}
\end{figure}

Similarly, we have done the same test for the real O3 noise and the simulated O4 Gaussian noise. The different tests are summarised in Fig.~\ref{fig:TAPcomp}, where each curve represents the results for the whole test set. In terms of PISNR, the efficiencies for O3 Gaussian noise and O4 Gaussian noise are very similar. However, since the noise floor is lower in the O4 case, the network can probe higher distances in this case. The performance for real noise is a bit worse than for the two Gaussian cases. The network needs a slightly larger PISRN to achieve the same performance. For example, the network needs a PISNR of 20 to have an efficiency of $50\%$ in the case of real O3, whereas it only needs a PISNR of 17 to reach the same sensitivity in the two other cases. Even if some glitches and non-Gaussian features are present in the data, the network is still able to reach a high performance provided that the PISNR is high enough. To be more realistic with a real time search, the noise is taken for the low-latency scenario. Therefore, it has a low quality, explaining the reduced performance.

After testing the network on independent 300 seconds-long frames, we generate longer frames of 1000 seconds, and we inject a complete GW signal into them. Then, we slide a 300 seconds window over the frame, pass the data in the window to the CNN for each step and make a prediction. From one step to the next, the window is shifted by 5 seconds. This is repeated until the full 1000 seconds are covered. Note that the step of 5 seconds is arbitrary and can be reduced, since for a realistic early-alert pipeline the length of the minimum step should be equal to the time required to load 300 seconds of data, pre-process it, and predict it with our network. The deep-learning algorithm is fast and takes about $0.5$~s on a CPU and $0.01$ s on a \textit{GeForce GTX 750 GPU}, the pre-processing is also fast: about $0.13$ s to compute the PSD with Pycbc, $0.01$ s to perform the whitening and $ 1$ s to remove the peaks and do the renormalization. The limiting factor is to load 300 seconds of data for 3 detectors with \textit{GWpy}, which takes around $2$ s on the LIGO servers. Note that the PSD used for the whitening is computed each time we load the $300$s frame. To reduce further this time, one can compute the PSD at regular intervals and use the result for multiple steps.

\begin{figure}[ht]
\begin{center}
    \includegraphics[scale=0.4]{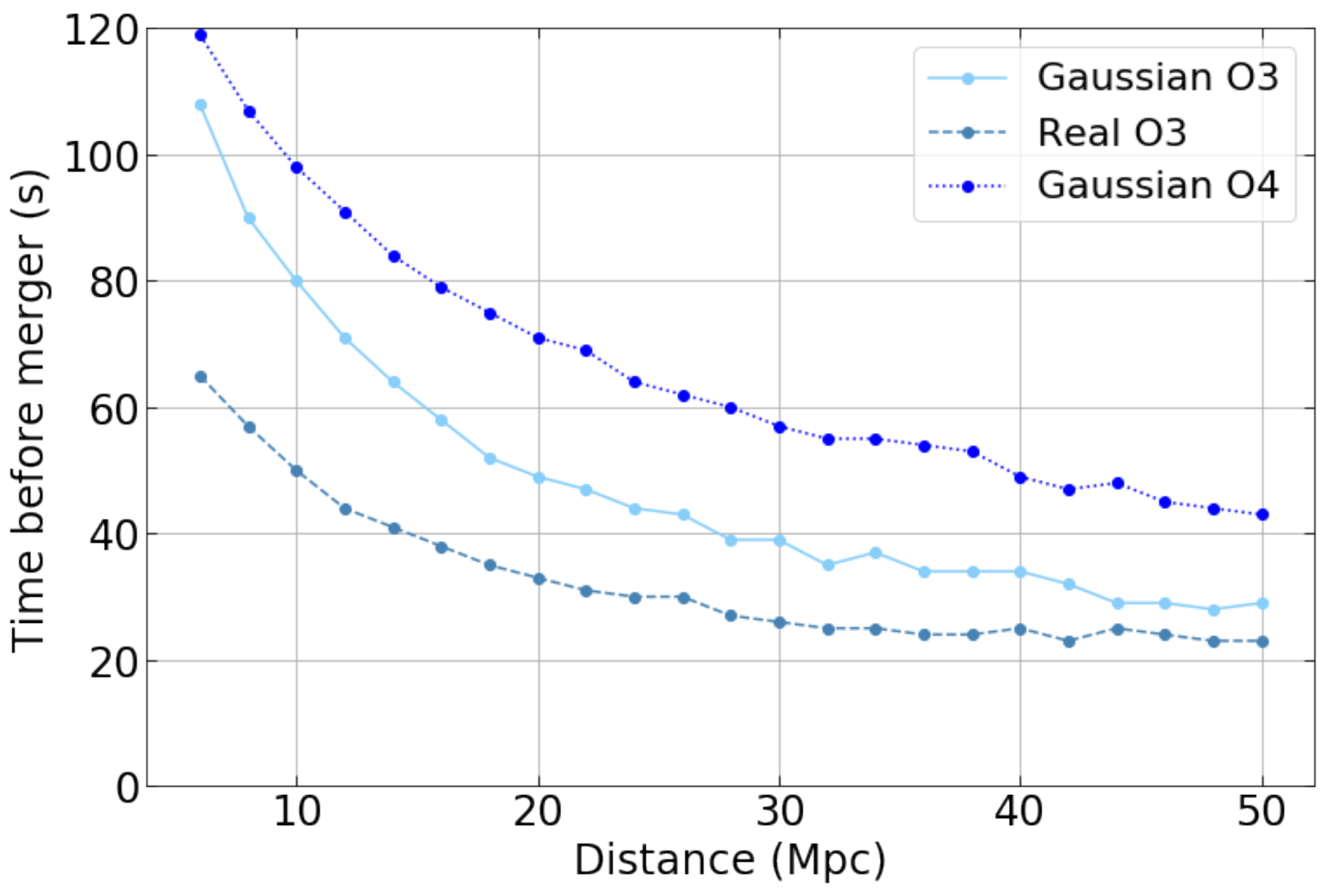}
    \caption{\label{fig:TBM_comp} The time before merger at which the event is detected as a function of the distance. These curves are made for a BNS with component masses similar to those of GW170817.}
\end{center}
\end{figure}

Fig.~\ref{fig:TBM_comp} illustrates the time left before merger when our approach is able to detect the event for the different noise types. Each point contains 1000 frames with a duration of 1000 seconds and each frame has a different noise realisation. In each frame, we inject a BNS signal with component masses similar to those detected for GW170817~\cite{abbott2017gw170817}. We choose fixed masses to keep the total duration of the signal fixed. The sky position of the signal is changed for each frame. We then slide a 300 seconds window over the 1000 seconds as described above. The process is then repeated for injections corresponding to a larger distance. Fig.~\ref{fig:TBM_comp} shows that, for a given distance, the events are detected the earliest in O4 Gaussian noise. It is also interesting to note that the time before merger for real O3 noise and Gaussian O3 noise are similar, even if the Gaussian case is slightly better.  An event like GW170817 at a distance of 40 Mpc can be detected by our method 25 s in advance in real O3 noise, 35 s in advance in Gaussian O3 noise, and 50 s in advance for Gaussian O4 noise, showing quite good trigger capabilities in future observations runs.

For an online matched filtering search, the performance is often evaluated by a false alarm rate (FAR). It represents the probability that a trigger occurs because of the noise for a given period of time~\cite{adams2016low}. We also define a FAR adapted to our method, as follows.

\begin{figure}[ht!]
\begin{center}
    \includegraphics[scale=0.4]{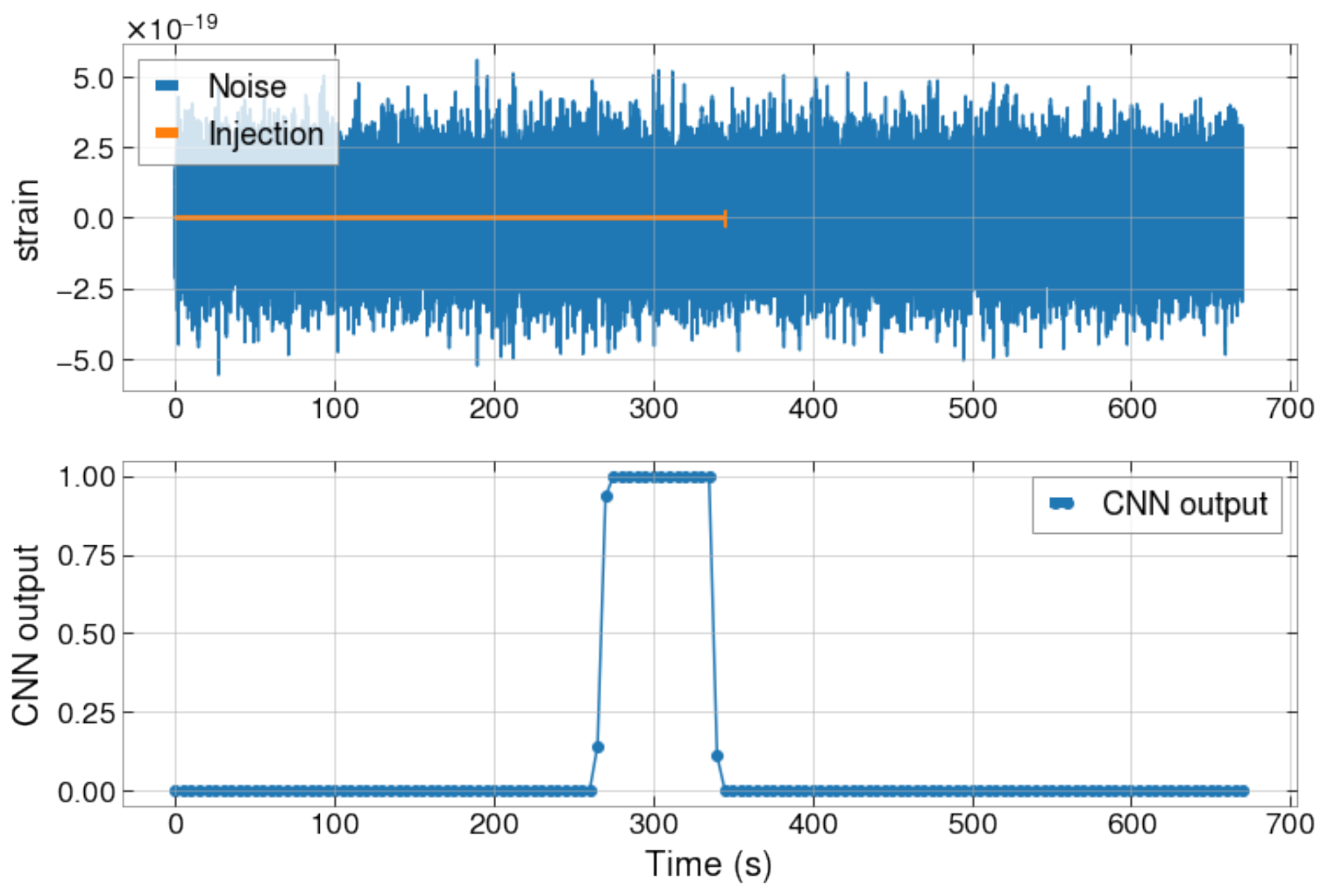}
    \caption{\label{fig:CNN_output} \textit{Top: } Representation of a signal and the noise it is injected in. \textit{Bottom: } Representation of the output of the CNN. Each point represents the probability to have an inspiral in the 300 seconds of data. By convention, the time of a point represents the end of the time window. The network does not trigger on the early inspiral because the PISNR is too low. When it becomes high enough, the networks produces a trigger until the injection leaves the frame, giving multiple points with a high probability in a row.}
\end{center}
\end{figure}

We run our network over the entire O3b data using the same setup as the one described previously. We shift the observation windows by 5 seconds for each step, and veto the times corresponding to events reported in the GWTC-3 catalog~\cite{LIGOScientific:2021djp} and assume that there are no other detectable events in the data\footnote{This assumption is reasonable since our network needs relatively high SNRs to detect the inspiral, and the event would therefore have been detected.}. The FAR is then defined as the number of triggers divided by the total observation time. For O3b, we obtain a FAR of 277.54 per day, which is too high to be used for online searches. To decrease its values, we can consider that an event is present when our network gives multiple triggers in a row, as shown in Fig.~\ref{fig:CNN_output}. If we keep detections with 5 consecutive triggers, the FAR goes down to 12.31 per day, and it goes to 1.71 per day if we consider 10 triggers in a row. The use of multiple triggers implies larger waiting times before producing an alert, and reduces the time before merger for the detection. For example, considering 5 triggers leads to a delay of 20 seconds as we consider steps of 5 seconds when sliding the window. In the end, this shows that we would need to find a trade-off between the time before merger and the desired FAR.

Another way to decrease the FAR is to use coherent triggers between two or more detectors. The training strategy for the network does not make it favor coherent triggers.  Indeed, since it is trained for one, two, or three detectors available, it learns to trigger even if only one detector is online. Furthermore, even if more than one detector is online, we do not use a minimum SNR in each detector for the training set. Hence the network learns to trigger even if only one interferometer is picking up the signal. In the end, this means that as soon as the CNN sees something remotely close to a signal in one of the detectors, it triggers, leading to a relatively high FAR.

\subsection{Number of BNS inspirals detectable in O4}
\label{Subsec:O4BNS}

\begin{table*}[t!]
    \centering
    \begin{tabular}{ l c c c c c   }
     \hline
     \hline
     Time before merger  & $\mathcal{M}_c$ $(M_{\odot})$ & network SNR & net PISNR at detection & Maximum frequency\\
     \hline
     \hline
     88 s & 1.19 & 71.87 & 15.32 & 25.45 Hz \\
     59 s & 1.08 & 63.77 & 23.43 & 31.35 Hz \\
     58 s& 1.26 & 53.01 & 16.63 & 28.75 Hz  \\
     25 s & 1.16 & 28.72 & 16.31 & 41.39 Hz \\
     22 s & 1.95 & 64.07 & 19.8 & 31.45 Hz \\
     22 s & 2.06 & 54.88 & 18.01 & 30.42 Hz \\
     19 s & 2.15 & 30.55 & 10.37 & 31.29 Hz\\
     14 s & 1.69 & 31.26 & 14.42 & 40.75 Hz \\
     11 s & 1.98 & 27.68 & 13.63 & 40.43 Hz \\
     10 s & 2.0 & 28.95 & 16.28 & 41.58 Hz \\
     10 s & 1.79 & 25.04 & 12.9 & 44.52 Hz \\
     7 s & 2.01 & 20.47 & 11.87 & 47.48 Hz \\
     7 s & 1.72 & 34.21 & 25.71 & 52.2 Hz \\
     3 s & 2.12 & 28.28 & 20.68 & 63.01 Hz \\
     \hline
    \end{tabular}
    \caption{\label{tab:Realpop} The time before merger, the maximum frequency seen by the network at detection, the chirp mass of the event, the network SNR, and the PISNR at the moment of the detection for all the detected BNSs in five years of simulated O4 data.}
\end{table*}

To estimate the number of BNSs that our network could detect in O4, we simulate a population of BNSs. It is generated using the method described in~\cite{Samajdar:2021egv} and the BNS merger rate is normalised so that the local rate is equal to the median rate given in~\cite{LIGOScientific:2020kqk}. The only difference with ~\cite{Samajdar:2021egv} is that we adapt the detection thresholds and the PSDs to our O4 scenario. We keep the BNS events with a network SNR higher than 13 and discard all the others. This threshold is chosen as we expect our network to find only BNSs that are clearly visible in the detector network and a global SNR of 13 corresponds approximately to an SNR of 8 in each detector. 

\begin{figure}
\begin{center}
    \includegraphics[scale=0.4]{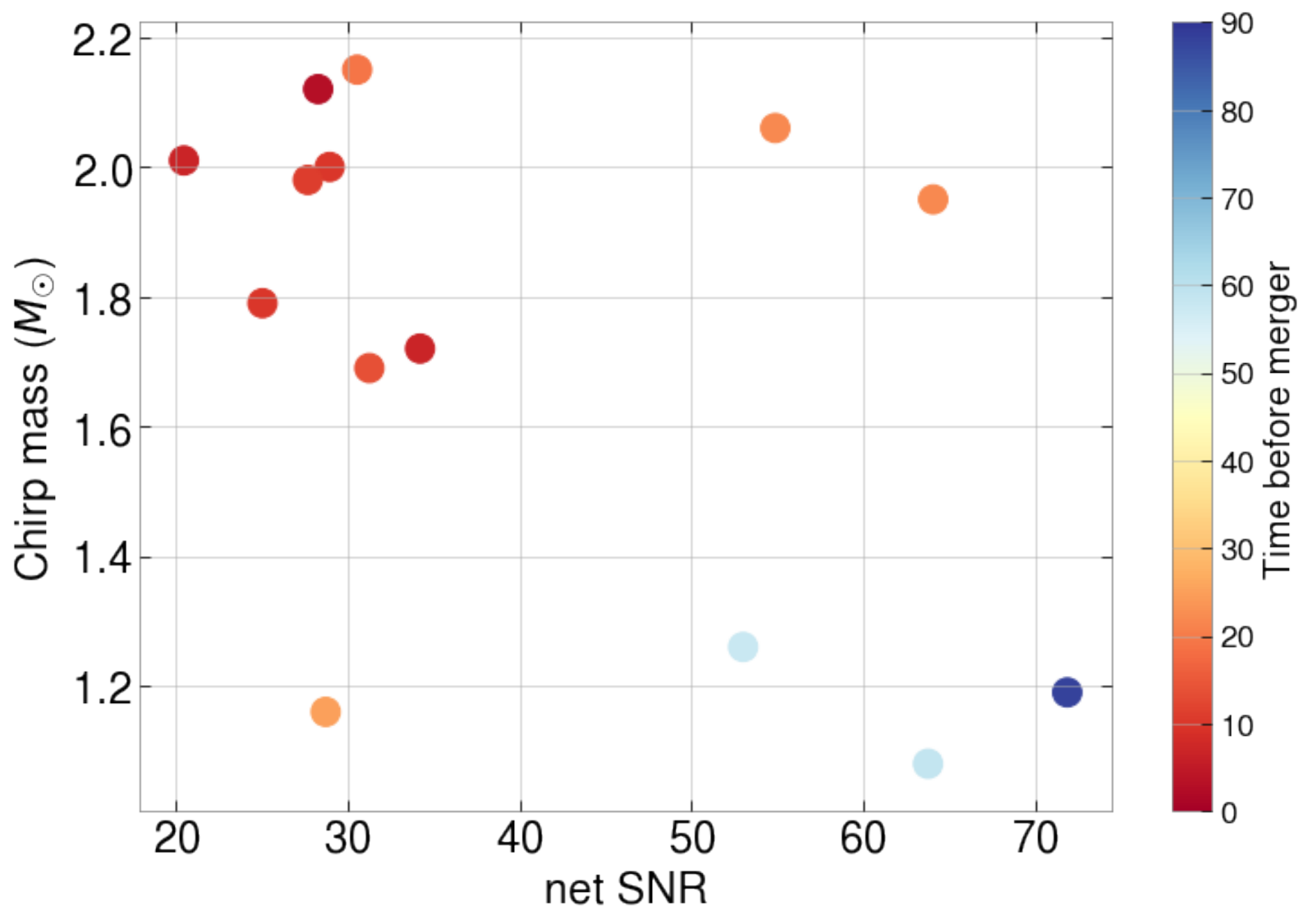}
    \caption{\label{fig:Real_pop} The number of BNSs detected in advance by our network for a simulated population of BNSs in five years of O4 data.}
\end{center}
\end{figure}

To have more statistics, we compute the equivalent of 5 years of data, and we consider a duty cycle of $100 \%$ for all the detectors. Our simulations predict that, on average, around twenty BNSs per year will have a network SNR over $13$, and our network can detect around three of those BNSs in advance. Fig.~\ref{fig:Real_pop} represents the time before the merger of all the BNSs detected by our network for the 5 years of generated data. Even if our network is able to detected only three events out of twenty, it is important to note that these events are seen in advance and would therefore not be seen at that stage by the unmodified matched filtering searches. Nevertheless, matched filtering pipelines adapted to the early detection of long inspirals are also being developed \cite{nitz2020gravitational, sachdev2020early}. Those are also able to detect BNS mergers in advance. Even if the comparison between these works and ours is difficult (partially because of the difference in noise, but also in performance evalution), we can mention that times before merger of these algorithms are comparable to those obtained by our network, ranging from $\mathcal{O}(10)$  to  $\mathcal{O}(100)$  seconds. An advantage of these early-warning matched filtering searches is that their FAR is lower than ours (around one per month) but they require more computational resources than our approach. 

Table~\ref{tab:Realpop} shows the different characteristics of the detected BNSs. The network can see an event when the net PISNR is between 10 and 25, which is expected according to Fig.~\ref{fig:TAPcomp}. The time before the merger at which the CNN can detect a signal  depends on two factors: a) the network PISNR and b) the length of the signal. The PISNR can be seen as a fraction of the SNR and its exact value depends on which part of the signal we are considering (hence the maximum frequency see by the CNN). Therefore, for a fixed signal duration, if the network SNR is high, the network can detect an event at a lower maximum frequency, corresponding to a longer duration before the merger. However, if we fix the SNR and the maximum frequency while increasing the duration of the signal (for example by decreasing both the chirp mass and the luminosity distance to compensate), the event will be detected with a larger duration before the merger. This behavior is well represented in Fig.~\ref{fig:Real_pop}, where events with a light chirp mass and a high SNR are detected the earliest. It also explains why some events with a lower chirp mass can be detected earlier, even if the network SNR is smaller than for other events.
\section{Conclusion}
\label{Sec:Conclusion}
This work builds upon the framework developed in~\cite{baltus2021convolutional}. We implement several upgrades and modifications to the CNN-based pipeline designed to detect the early inspiral phase of BNS events. A major upgrade is the increased duration of the frames passed to the network. That allows us to search for smaller frequencies and opens the door to earlier detections. Another benefit of this increased duration is that we can use a single network to look for all type of BNSs, which was not the case in our previous work. The detection of events for a smaller maximum frequency is not easy and required an adapted training methodology: curriculum learning. We consider realistic observation scenarios, including all the detectors of the LIGO-Virgo network and use realistic noise realisations: O3 and O4 Gaussian noises, as well as real O3 low-latency noise. We have also demonstrated that even in the real O3 noise our network is able to detect GW signals in advance. We expect our network to detect some BNSs in O4, up to minutes in advance if the SNR of the event is high enough. In  future works, we will upgrade our method to search for neutron-star-black-hole mergers as well. As discussed in section \ref{Sec:Results}, we will also develop methods to decrease the FAR. Finally, we will investigate a way to infer the sky position with only the early inspiral part.

\section*{Acknowledgments}
The authors thank Thomas Dent and Srashti Goyal for their useful comments, as well as Maxime Fays, Vincent Boudart, Sarah Caudill and Chris Van Den Broeck for useful discussions. G.B. is supported by a FRIA grant from the Fonds de la Recherche Scientifique-FNRS, Belgium. J.R.C. acknowledges the support of the Fonds de la Recherche Scientifique-FNRS, Belgium, under grant No. 4.4501.19. M.L. H.N., and J.J  are supported by the research program of the Netherlands Organisation for Scientific Research (NWO). The authors are grateful for computational resources provided by the LIGO Laboratory and supported by the National Science Foundation Grants No. PHY-0757058 and No. PHY-0823459. This material is based upon work supported by NSF's LIGO Laboratory which is a major facility fully funded by the National Science Foundation.

\appendix

\onecolumngrid

\section{Details on the veto of peaks for real noise}
\label{appendix:1}

\begin{figure}[h]

\begin{center} 
    \includegraphics[scale=0.45]{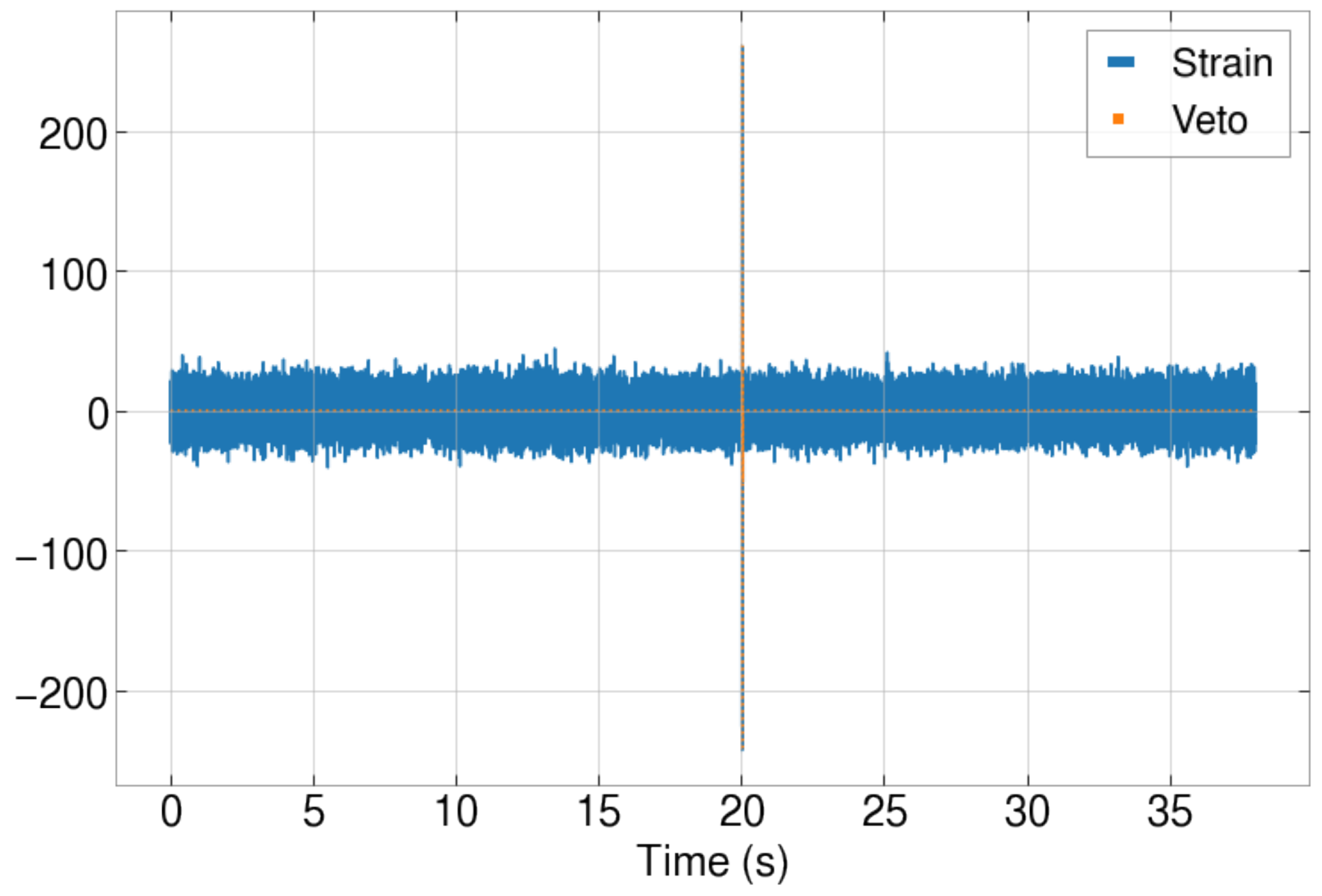}
    \caption{\label{fig:peak} The blue curve represents O3a noise after application of the whitening, the low-pass filter, and the high-pass filter. The orange curve shows the part which will be vetoed.}
\end{center}
\end{figure}

After the noise has been downloaded and whitened, large peaks can still be present in the data (see Fig.~\ref{fig:peak}). This behavior only appears for real O3 noise, and leads to a problem for the normalisation. Indeed, before passing the data to the network, we normalise them to be between -1 and 1. To do so, we find the maximum absolute value of the strain and divided each point by that value. When a large peak is present, the maximum absolute value is the value of the peak and it makes the rest of the time series too small. That confuses the neural network, and we decided to veto these peaks. The vetoing is done according to the z-score, which is defined as:
\begin{equation}
    \label{zscore}
    Z_{i} = \frac{x_{i} - \mu}{\sigma}
\end{equation}
where $x_{i}$ is the value of a point $i$ in the time series, $\mu$ and $\sigma$ are respectively the mean and the variance of the time series.
We then compute the standard deviation of the z-score and put all the points with a z-score larger than 5 times the standard deviation to zero, allowing to remove large peaks such as those seen in Fig.~\ref{fig:peak}. The normalisation is then done on the vetoed frame.

\twocolumngrid

\bibliographystyle{apsrev}
\bibliography{Manuscript}{}

\end{document}